\documentclass[apl, twocolumn,tightenlines]{revtex4-1}

\usepackage{amsthm}
\usepackage{amsmath}
\usepackage{bbm, dsfont}
\usepackage{graphicx}
\usepackage[usenames,dvipsnames]{color}
\usepackage[colorlinks=true,citecolor=magenta,urlcolor=blue]{hyperref}
\usepackage[table]{xcolor}
\usepackage{colortbl}
\usepackage{color}
\usepackage{multirow}
\usepackage{hhline}
\usepackage{tabu}
\usepackage{epstopdf}
\usepackage{booktabs}
\usepackage[normalem]{ulem}
\usepackage{siunitx}
\usepackage{enumitem}
\usepackage{braket}

\newcolumntype{C}[1]{ >{ \centering\arraybackslash}p{#1}}

\newcommand{\X}{$\mathsf{X}$}
\newcommand{\Z}{$\mathsf{Z}$}

\newcommand{\pax}{p_\mathsf{X}^\text{A}}
\newcommand{\paz}{p_\mathsf{Z}^\text{A}}
\newcommand{\pbx}{p_\mathsf{X}^\text{B}}
\newcommand{\pbz}{p_\mathsf{Z}^\text{B}}

\newcommand{\nz}{n_\mathsf{Z}}

\newcommand{\QZ}{Q_\mathsf{Z}}

\newcommand{\epscorr}{\epsilon_\text{corr}}
\newcommand{\epssec}{\epsilon_\text{sec}}

\newcommand{\sz}[1]{s_{\mathsf{Z}, #1}}

\newcommand{\phiz}{\phi_{\mathsf{Z}}}

\begin{document}
\title{Simple and high-speed polarization-based QKD}

\author{Fadri Gr\"unenfelder}
\affiliation{Group of Applied Physics, University of Geneva, Chemin de Pinchat 22, CH-1211 Geneva 4, Switzerland}
\author{Alberto Boaron}\email{alberto.boaron@unige.ch}
\affiliation{Group of Applied Physics, University of Geneva, Chemin de Pinchat 22, CH-1211 Geneva 4, Switzerland}
\author{Davide Rusca}
\affiliation{Group of Applied Physics, University of Geneva, Chemin de Pinchat 22, CH-1211 Geneva 4, Switzerland}
\author{Anthony Martin}
\affiliation{Group of Applied Physics, University of Geneva, Chemin de Pinchat 22, CH-1211 Geneva 4, Switzerland}
\author{Hugo Zbinden}
\affiliation{Group of Applied Physics, University of Geneva, Chemin de Pinchat 22, CH-1211 Geneva 4, Switzerland}

\begin{abstract}
We present a simplified BB84 protocol with only three quantum states and one decoy-state level. We implement this scheme using the polarization degree of freedom at telecom wavelength. Only one pulsed laser is used in order to reduce possible side-channel attacks. The repetition rate of 625\,MHz and the achieved secret bit rate of 23\,bps over 200\,km of standard fiber are the actual state of the art.
\end{abstract}

\maketitle

 
Secure communication is a central pillar of today's society, playing a key role not only in finance, defence and industry, but also in the protection of the privacy of individuals. 
Most cryptographic systems used at present, however, are lacking an information theoretical security proof and are threatened by future quantum computers. 
Quantum key distribution (QKD) offers a way to overcome this security issue by exchanging a secret key over an insecure optical link. 
This key can be used in the One-Time-Pad for secure communication~\cite{Lo2014a}. 

The idea of QKD was born in 1984 when Bennett and Brassard proposed a protocol which is now known as BB84 \cite{Bennett1984}. 
Nowadays, a variety of different protocols exists and many implementation using different degrees of freedom (DoF) of photons (polarization, phase, etc.) have been demonstrated using optical fibers or free-space~\cite{Ekert1991,Bennett1992,Grosshans2001,Inoue2002,Stucki2005,Sasaki2014}. 

For practical reasons, implementations of polarization-based BB84 often use weak pulses from several different laser diodes, one for each qubit state~\cite{Bienfang2004,Tang2006,Liu2010,Liao2017}. However, different lasers may have slightly different properties such as frequency and emission time, offering to an eavesdropper Eve the possibility of a so-called side-channel attack. Eve, by looking at those properties, may determine the qubit states sent without disturbing them~\cite{Nauerth2009,Sun2017}.

In this paper, we present a complete polarization-based QKD setup based on a single laser in order to prevent side-channel attacks exploiting the distinguishability of different lasers.
The source works at a repetition rate of \SI{625}{\mega\hertz}.
We reduce the complexity of the scheme as much as possible, using a three-state protocol~\cite{Boaron2016}, only one decoy-state level~\cite{Ma2005,Lim2014} and only two single-photon detectors.  We perform a complete key exchange, with real-time error correction and privacy amplification based on finite-key analysis.

\vspace{0.4cm}

In the following, we describe the protocol step by step.

\noindent \textbf{1. State preparation:}
Alice uses a laser source emitting phase-randomized weak coherent pulses and encodes the states in the polarization DoF of the photons.
She chooses randomly one of the two bases \X{} or \Z{} with the associated probabilities $\pax$ and $\paz = 1-\pax$, respectively. 
In the basis \Z{}, she generates with uniform probability the state $\ket{H}$ or  $\ket{V}$.
In the basis \X{}, she just prepares $\ket{+}=(\ket{H}+\ket{V})/\sqrt{2}$.
The pulse energy is chosen at random among one of the two mean photon numbers $\mu_1$ and $\mu_2$  with the constraint $\mu_1 > \mu_2 > 0$, and probabilities $p_{\mu_1}$ and $p_{\mu_2} = 1 - p_{\mu_1}$, respectively.
$\mu_1$ is denoted as signal level while $\mu_2$ is the decoy level.
Alice sends the qubits through an optical fiber to Bob. 

\noindent \textbf{2. Measurement:}
Bob performs a measurement on the incoming signal at random in one of the two bases \X{} or \Z{} with respective probabilities $\pbx$ and $\pbz = 1- \pbx$.
For each detection, the basis and the measurement result are recorded.  

\noindent \textbf{3. Basis reconciliation:}
Alice and Bob announce their basis settings for the events where a detection has occurred. 
The events from the \Z{} basis are used to generate the raw key, while those from the \X{} basis are used to estimate the eavesdropper potential information.
After having collected $\nz^\text{EC}$ new raw key bits, they continue with step 4.

\noindent \textbf{4. Error correction:}
Alice and Bob apply an error correction algorithm on the block of  $\nz^\text{EC}$ bits during which $\lambda_{\text{EC}} = f_{\rm{EC}} \cdot \nz^\text{EC} \cdot h(\QZ)$  bits are disclosed where $f_{\rm{EC}}$ is the efficiency of reconciliation, $h(x)$ the binary entropy, and $\QZ$ the error rate. 
In our protocol, we employed the error correction algorithm Cascade which has a reconciliation efficiency around 1.06 \cite{Martinez2015}.
The procedure succeeds with a probability $1-\epscorr$.
After $k = \nz/\nz^\text{EC}$ error correction blocks they proceed to step 5.

\begin{figure*}[t]
\includegraphics[width = 1.9\columnwidth]{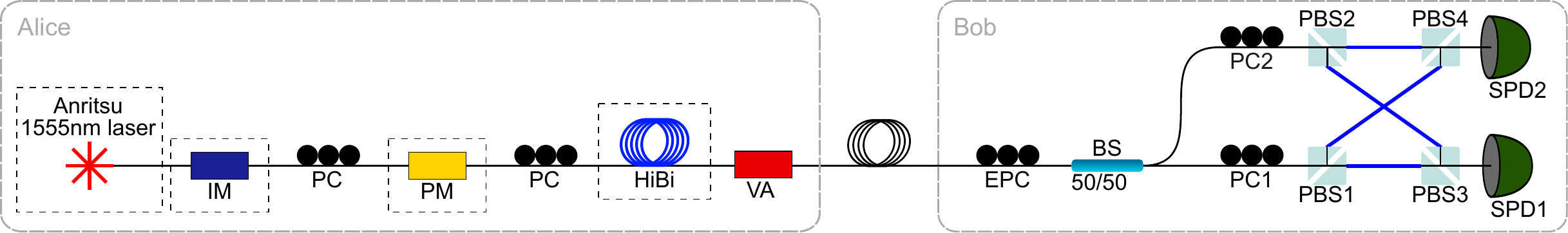}
\caption{\label{fig:Setup} Schematics of the experimental setup. IM: intensity modulator; PC: polarization controller; PM: phase modulator based on a lithium niobate waveguide; HiBi: high birefringence fiber; VA: variable attenuator; PBS: polarizing beamsplitter; BS: beamsplitter; EPC: electronic polarization controller; SDP: single-photon detector. 
The arm connecting the BS and the PC2 introduces a delay of \SI{800}{\pico\second} compared to the arm from the BS to the PC1 due to a difference in length. 
The blue lines connecting the PBS are polarization-maintaining fiber and have all the same length.
Dashed lines denote temperature stabilized boxes.}
\end{figure*}

\noindent \textbf{5. Privacy amplification:}
Alice and Bob apply the privacy amplification procedure on a block of size $\nz$ to obtain a secret key of $l$ bits~\cite{Bennett1995}. 
$l$ is upper bounded by:
\begin{align}\label{eq:skl}
l \leq &  \sz{0} + \sz{1}(1-h(\phiz)) - \lambda^t_\text{EC} \nonumber\\
& - 4\log_2(7/\epsilon_\text{sec}) - \log_2(1/\epsilon_\text{cor}),
\end{align}
where $\sz{0}$ and $\sz{1}$ are the lower bound on the number of vacuum and single-photon detection in the \Z{} basis, $\phiz$ is the upper bound on the phase error rate, $\lambda^t_\text{EC}$ is the total number of bits revealed during the error correction, and $\epsilon_\text{sec}$ and $\epsilon_\text{cor}$ are the secrecy and correctness parameters, respectively.

Having only two intensity levels does not allow us to directly measure an upper bound on $\sz{0}$, which is necessary to estimate the lower bound on the single-photon events.
To solve this issue, we consider that the total number of errors on each basis is only due to the vacuum component. This is the most conservative way to estimate the upper bound of $\sz 0$ (for more details, see~\cite{Rusca2018}).

\vspace{0.4cm}

Now, we describe the experimental setup that is shown in \autoref{fig:Setup}.
The experiment is controlled by two field programmable gate arrays (FPGAs) placed on Alice's and Bob's side.
On Alice's side, a gain-switched DFB laser at \SI{1554.94}{\nano\meter} [Anritsu], triggered at \SI{625}{\mega\hertz}, generates phase-randomized weak coherent pulses with a pulse duration of \SI{93}{\pico\second}. 
An intensity modulator [Photline] based on an integrated Mach-Zehnder interferometer on lithium-niobate (LiNbO$_3$) then encodes the decoy levels. 
The polarization encoding is done by a titanium indiffused LiNbO$_3$ phase modulator (PM) [Thorlabs]. 
The pulses are injected in it with a polarization $(\ket{H}+\ket{V})/\sqrt 2$. 
We control the relative phase $\phi$ between $\ket{H}$ and $\ket{V}$ by applying a voltage on the PM to change its birefringence.
At the output we have then the state $(\ket{H}+e^{i\phi}\ket{V})/\sqrt2$. 
The phase $\phi \in \{0,\pi/2, \pi \}$ is randomly chosen by the FPGA and set by a 3-level digital-to-analog converter made in-house. 
The PM introduces a polarization mode delay of 10.7 ps that is compensated by 8 m of high birefringence (HiBi) fiber. 
Alice chooses the basis \X{} with a probability $\pax = \frac{1}{8}$.
At the output of Alice's device, an attenuator sets the correct mean photon number of the outgoing pulses.

Bob's basis choice is made by a symmetric beamsplitter (BS), meaning that $\pbz = \pbx = \frac{1}{2}$.
The splitting ratio could be optimized for each distance.
However, for short distances $\pbz>\frac{1}{2}$ is not advantageous since the detectors are saturated, and for longer distances $\pbx$ has to be increased to have enough statistics in the \X{} basis.
So for simplicity we chose a 50:50 splitting ratio as a good compromise for almost all distances.
Two polarization controllers, PC1 and PC2, are set such that the two fiber-based polarizing beamsplitters, PBS1 and PBS2 (extinction ratios $> 20\,\rm{dB}$), perform a projection in the \Z{} (rectilinear) and \X{} (diagonal) basis, respectively.
The output ports corresponding to $\ket{H}$ and $\ket{V}$ are recombined with $\ket{+}$ and $\ket{-}$ via two other PBSs. 
To distinguish between the two bases, an additional delay of 800~ps is introduced in the arm of the \X{} basis. 
With this temporal multiplexing we are able to use only two detectors instead of four.
We employ in-house made free-running single-photon detectors based on InGaAs/InP negative feedback avalanche photodiodes cooled by a free-piston Stirling cooler to achieve dark counts rates of 10\,Hz~\cite{Korzh2014}.
Note that for shorter distances up to 100\,km it would be more appropriate to use four detectors in order to reduce the saturation.
Moreover, Peltier cooling would be sufficient, as dark counts are less critical.

The polarization of the pulses during transmission is prone to fluctuations, e.g. due to temperature drifts. 
To compensate for these fluctuations, we have implemented a feedback loop based on trial-and-error approach that acts on an electronic polarization controller (EPC) [Phoenix Photonics] placed at the input of Bob's setup.
This EPC is composed of three adjustable phase plates based on small HiBi fiber pieces whose temperature is adjusted to change their birefringence.
The third one is set such that it affects only the phase between $\ket{H}$ and $\ket{V}$, and by consequence the quantum bit error rate (QBER) in the \X{} basis. 
Thus, the feedback loop takes the QBER in the \Z{} basis as error signal to control the first two wave-plates and in the \X{} basis for the third one.
Note that, the QBER in the \X{} basis is directly given by the probability to detect the state $\ket{-}$ when the basis \X{} is prepared. 
In the \Z{} basis, the QBER $\QZ$ is provided in real-time by the Cascade error correction algorithm.
This approach exempts us to use additional lasers to monitor the polarization drifts~\cite{Xavier2008}.

\vspace{0.4cm}

We perform exchanges of secret keys with complete distillation, i.e. taking into account the finite statistics effect for the privacy amplification, for different transmission distances.
The quantum channel is composed of a fiber spool of 12\,km and a variable attenuator set to a value $\eta_{\rm{att}}$ to simulate additional optical fiber with a loss of 0.2\,dB/km.
In order to test the polarization stabilization scheme, we also perform a key exchange with more than 100\,km of real fiber.
For all measurements, the sizes of the error correction blocks and the privacy amplification blocks are set to $\nz^\text{EC} = 8192$ and $\nz = \num{8.192e6}$, respectively.
These parameters offer a good compromise between the time of acquisition and the effect of finite-key statistics on the secret key rate (SKR).
The security parameters are fixed to $\epssec = 10^{-9}$ and $\epscorr = 10^{-15}$.

\begin{figure}
\includegraphics[ trim={0.2cm 0 0 0},clip, width = \columnwidth]{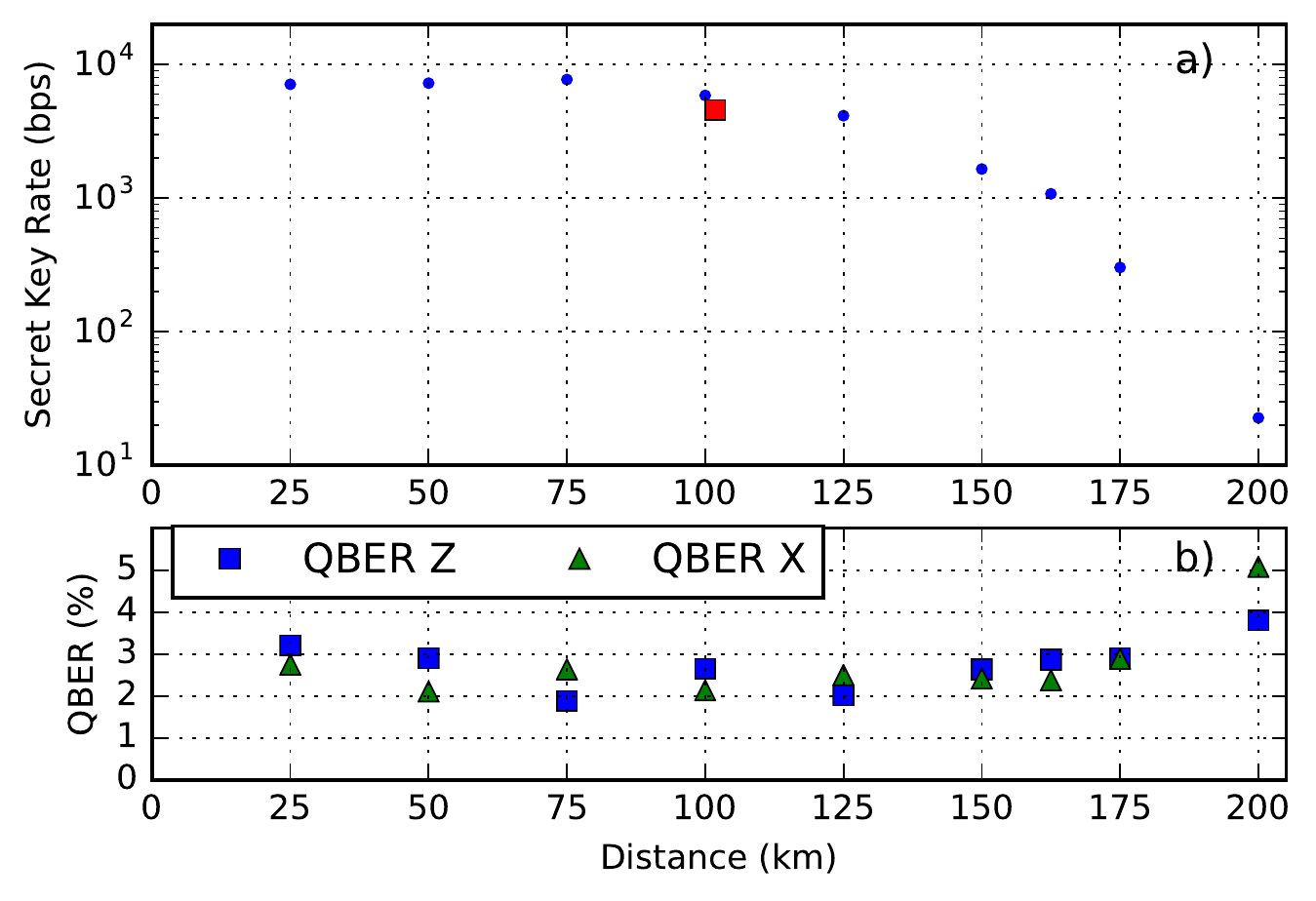}
\caption{\label{fig:skr} SKR (a) and QBER (b) as a function of the transmission distance. The measurements were done using a quantum channel composed of $\SI{12.16}{\kilo\meter}$ of real fiber and attenuation to simulate additional distance. The red square indicates a measurement done with more than \SI{100}{\kilo\meter} of real fiber.}
\end{figure}

For every result depicted in \autoref{fig:skr}, the SKR has been maximized by optimizing the mean photon numbers $\mu_1$, and $\mu_2$, the probability $p_{\mu_1}$ and the detectors parameters i.e. temperature, dead-time and efficiency.
The QBER due to polarization misalignment is automatically minimized via the feedback that acts on the EPC.

From \autoref{fig:skr}, we can clearly identify three regimes.
Up to 125 km, the SKR is mainly limited by the saturation of the detectors due to a dead-time around 30\,$\mu$s.
Therefore it is favourable to keep the mean photon number of the states sent by Alice low. At 25~km we use $\mu_1 = 0.10$, $\mu_2 = 0.06$ and $p_{\mu_1} = 0.56$.
In this range the SKR could be improved by employing faster single-photon detectors such as superconducting
nanowire single-photon detectors~\cite{Wang2012b} or gated avalanche photodiodes~\cite{Comandar2014}.
Nevertheless, the InGaAs single-photon detectors we use are much less complex.
Above 125\,km, the SKR decreases exponentially as expected due to the fiber loss, until around 175\,km where the dark-count rate becomes significant compared to the detection rate and as a consequence the QBER increases rapidly.
At this distance, the settings are $\mu_1 = 0.33$, $\mu_2 = 0.14$ and $p_{\mu_1} = 0.75$.
We achieve a SKR of 303\,bps.
This result is comparable to other state of the art of long-distance QKD experiments~\cite{Takesue2007b,Stucki2009a,Wang2012b,Korzh2015}.
Moreover, the SKRs are better than other two-decoy/four-state BB84 experiments \cite{Liu2010,Frohlich2017}.
Indeed, our simulations show that up to about 175~km, the one-decoy level approach is slightly more efficient than the two-decoy one.
Finally, the SKR at 200\,km is 23\,bps.

\vspace{0.4cm}

To conclude, we implemented a BB84 protocol with states encoded in the polarization DoF of weak coherent pulses.
Our source is based on only one pulsed laser in order to prevent side-channel attacks.
It could be used both for fiber and free space implementations.
We kept the system simple with a three-state encoding approach with only one decoy-state level.
Therefore we have in total 6 different states instead of 12 for the complete protocol, which greatly simplifies the state preparation and the data processing. 
Using a rigorous security analysis taking into account finite-key effects, we distilled secret keys at a rate of 23\,bps for a distance of 200 km.

\section*{Acknowledgements}
We would like to acknowledge Jes\'us Mart\'inez-Mateo for providing the error correction code, Gianluca Boso for his contribution in building the control electronics, Raphael Houlmann for the FPGA programming and Charles Ci Wen Lim for the useful discussions about the security proof. We thank the Swiss NCCR QSIT and Davide Rusca thanks the EUs H2020 programme under the Marie Sk\l{}odowska-Curie project QCALL (GA 675662)  for financial support.

\bibliography{bib_bb84}
\end{document}